# A compact and versatile microfluidic probe for local processing of tissue sections and biological specimens

*J.F. Cors, R.D. Lovchik, E. Delamarche and G.V. Kaigala**

IBM Research—Zurich

Saeumerstrasse 4, CH-8803 Rueschlikon,

Switzerland.

E-mail: gov@zurich.ibm.com

**ABSTRACT**

The microfluidic probe (MFP) is a non-contact, scanning microfluidic technology for local (bio)chemical processing of surfaces based on hydrodynamically confining nanoliter volumes of liquids over tens of micrometers. We present here a compact MFP (cMFP) that can be used on a standard inverted microscope and assist in the local processing of tissue sections and biological specimens. The cMFP has a footprint of $175 \times 100 \times 140$ mm$^3$ and can scan an area of $45 \times 45$ mm$^2$ on a surface with an accuracy of $\pm 15$ $\mu$m. The cMFP is compatible with standard surfaces used in life science laboratories such as microscope slides and Petri dishes. For ease of use, we developed self-aligned mounted MFP heads with standardized "chip-to-world" and "chip-to-platform" interfaces. Switching the processing liquid in the flow confinement is performed within 90 seconds using a selector valve with a dead-volume of approximately 5 $\mu$L. We further implemented height-compensation that allows a cMFP head to follow non-planar surfaces common in tissue and cellular ensembles. This was shown by patterning different macroscopic copper-coated topographies with height differences up to 750 $\mu$m. To illustrate the applicability to tissue processing, 5 $\mu$m thick M000921 BRAF V600E+ melanoma cell blocks were stained with hematoxylin to create contours, lines, spots, gradients of the chemicals, and multiple spots over larger areas. The local staining was performed in an interactive manner using a joystick and a scripting module. The compactness, user-friendliness and functionality of the cMFP will enable it to be adapted as a standard tool in research, development and diagnostic laboratories, particularly for the interaction with tissues and cells.

## I. INTRODUCTION

The MFP is a non-contact, scanning microfluidic technology, which enables local (bio)chemical reactions on surfaces by hydrodynamically confining nanoliter volumes of reagents. This approach for localizing biochemical reactions on surfaces can benefit many areas, in particular in the life sciences. The MFP, on account of being able to perform localized chemistry within a liquid environment, has a strong potential for applications in tissue staining, electrophysiology, synaptic signaling, single cell transfection, chemical locomotion, and cell nanotoxicology, establishing conditions for protein crystallization, for example. In particular, there is a significant potential for use of such a technology in processing and interfacing locally with tissue sections and cellular ensembles for diagnostics and research. To catalyze the broader use of this technology, it is necessary to reduce the footprint of the current bulky MFP[1] and make it a user-friendly system. This will likely enable its routine use in standard "non-microfluidic" research and diagnostic laboratories akin a stand-alone tool. We here present a cMFP, which can be placed on the stage of a standard inverted microscope providing easy and rapid access to the MFP technology, when required, without the need of modifying or assigning a dedicated microscope. In addition, we present a strategy for patterning curved and irregular surfaces enabling interactions at the microscopic scale with macroscopic topographies.

Local surface (bio)chemistries in a non-contact mode can be performed with techniques such as dip pen nanolithography (DPN), nanopipettes and liquid encapsulation.[2] DPN uses ink diffusion between a cantilever tip of an atomic force microscope and a surface to deliver chemicals onto a substrate with sub-micrometer resolution.[3] This technique has been used for patterning biomolecular micro- and nanoarrays, tailoring chemical surfaces and the fabrication of metal and semi-conductor structures.[4] The precision of the control over the distance between the tip and the surface dictates the roughness and the elasticity of substrates that can be used. Additionally, the distance feedback may be confounded during operation in liquid environments and at lengthscales in the micrometer range. In contrast, nanopipettes deliver charged molecules onto a surface with an electric field between the pipette and the substrate.[4,5] They have been applied in several areas, for example, in the production of DNA microarrays, assembly of complex molecular structures and DNA trapping.[5] The use of thin glass structures (100 nm) however makes nanopipettes fragile and they also require conductive substrates. Aqueous two-phase systems (ATPSs), a more recent technique, enables the local delivery of liquids onto a surface by making use of compartmentalization.[6] This partitioning has been realized with dextran and polyethylene glycol.[7] ATPS was applied to many systems, including the patterning of one type of cells onto another cell type[8] and studying cell-to-cell contact.[9] This technique may be limited by the choice of the immiscible liquids and their biological compatibility. While the above techniques are well suited for specialized applications and scenarios, they are not readily suited for use with biological surfaces immersed in liquid together with the flexibility of localizing chemicals and scanning at the length-scales used in cell biology and tissue handling.

Hydrodynamic flow confinement (HFC) leverages the characteristics of laminar flows outside of a closed channel, in the "open-space".[10] To achieve HFC, a minimum of two channels is required, one for the injection of a processing liquid and the second one for aspiration. The MFP confines nanoliter volumes of processing liquid by hydrodynamically focusing it in the presence of an immersion liquid at length scales of tens of micrometers. The confinement is formed at the head, which is a silicon or silicon-glass hybrid device comprising micron-sized channels, apertures and an apex. The head is positioned a few micrometers above a surface. Head designs can vary from being coplanar to the scanned surfaces[11,12] or vertically oriented.[1,13] The MFP has been applied for protein patterning at the micrometer-scale, generation of chemical gradients, localized staining of cells, detachment of single cells[11,14] and perfusion of organotypic brain tissue.[15] More recently, the vertical MFP has even been applied to multiplexed immunohistochemistry on human tissue sections.[16]

Other groups made use of HFC, for example, Hitoshi *et al.* used a theta-shaped glass capillary to collect mRNA from adherent cells.[17] Perrault *et al.* developed a PDMS-based MFP to pattern fluorescein-labeled biotin on a streptadivin-coated slide.[18] The multifunctional pipette, a PDMS-based implementation with a three-channel arrangement was used for single-cell studies[19] and pharmacology.[20] We note that from an application standpoint these demonstrations are interesting, however the use of PDMS as the material to fabricate the devices, limits their use to aqueous solutions.[21]

Research on the MFP has mostly focused on selected applications in biology and medicine with a few exceptions, where fluid flow has been modeled[22] and on the instrumentation. For example, Juncker *et al.* developed a MFP system focusing on the integration at the software-level.[18] There still remain challenges in the integration at the hardware level. In this paper, we present a compact, modular and easy-to-use instrument by optimizing its design, the interfaces as well as the fluidic connections and illustrate its use in cell and tissue staining.

## II. SYSTEM CONFIGURATION OF THE cMFP

We designed the cMFP keeping in mind the need and relevance of a new tool that would significantly help in local processing tissue sections and biological specimens. The list of the requirements for such a technology is provided in Table 1. The cMFP has three main aspects, the positioning system, fluidics and the compatibility (Fig. 1).

**Table 1**. *Requirements imposed on a non-contact technology for local processing of tissues and cellular ensembles.*[23–25,26–31]

| | REQUIREMENTS | RELEVANCE |
|---|---|---|
| POSITIONING | Accuracy in the micrometer-range | Technologies aiming to work at the single-cell level (8-12 $\mu$m) require $\mu$m accuracy[18-20] |
| | Motorized control over the X-, Y- and Z-axis | Precise, repeatable and automated spatial positioning is key to any scanning technology[23] |
| | Mechanism for co-planarity between 'probe' and surface | Non-contact techniques require distance control mechanism and alignment of the probe with $\mu$m to nm precision, e.g. nanopipettes,[24-25] dip-pen nanolithography[4] and the microfluidic probe[1,11,18] |
| | Accurate head-to-surface distance control | |
| | Two modes of operation: continuous ($\mu$m/sec) and stop-and-go (mm/sec) | Continuous mode to make lines,[11,13] and stop-and-go to make spots[18,26] |
| FLUIDICS | Control of up to four syringe pumps | Hydrodynamic shaping of liquids needs individual control over multiple flows[12-27] |
| | Rapid liquid switching | In biological assays, sequential exposure to chemicals is required e.g. immunohistochemistry[28,29] |
| | Working volumes between 50 to 1000 $\mu$L | Typical volumes used in microfluidics[30,31] |
| COMPATIBILITY | Compatible with typical substrates in the life-sciences | Microscope slides (49 × 49 mm and 75 × 25 mm) and Petri dishes (Ø 90 mm)[2] with thickness up to 1 mm |
| | Visualization | |
| | Device footprint | Typical stage size of a standard inverted microscope is ~ 230 × 300 mm |

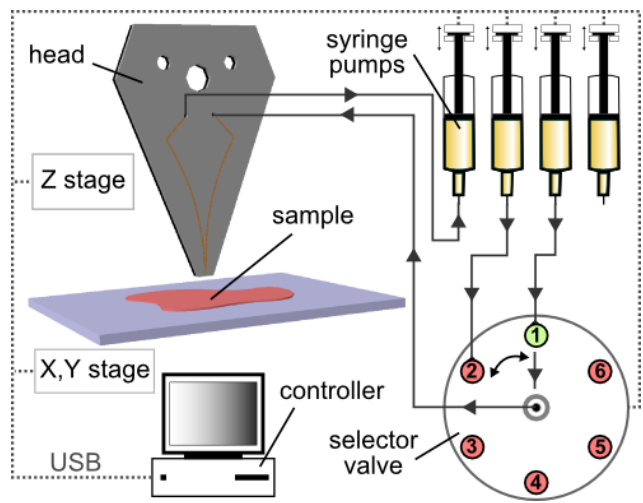

*FIG. 1. Modular architecture of the cMFP with each module being computer controlled.*

## A.  MFP heads

The MFP head is a microfabricated hybrid silicon-glass device with micrometer-sized features fabricated in-plane on the silicon layer.[1] The head in its simplest form has a "diamond" shape with 1 cm$^2$ surface area, an apex of 1 mm$^2$ and a thickness of 1 mm. These dimensions enable the formation of a flow confinement of approximately 0.006 mm$^2$. Two micrometer-sized channels with cross-sections ranging from $1 \times 1$ $\mu$m$^2$ to $50 \times 50$ $\mu$m$^2$ separated by 10 to 100 $\mu$m terminate at the apex. Here, we used a head with $50 \times 50$ $\mu$m$^2$ channels separated by 50 $\mu$m. The diamond shape allows scanning over a larger area ($45 \times 45$ mm$^2$) of a sample with the tapered end of the MFP head. Typically, a processing liquid is injected through one channel at a flow rate ($Q_i = 1-2$ $\mu$L/min), and is aspirated together with immersion liquid through the other channel ($Q_a = 5-10$ $\mu$L/min), resulting in the spatial hydrodynamic confinement of the processing liquid. During operation, the apex of the MFP head is placed 1 to 50 $\mu$m above the sample. Using these parameters allows performing chemical events on surfaces with nanoliters of reagents.

**B. Positioning system**

The positioning system is used to position the sample and the head, and to align the head with the objective of the microscope. The positioning system is made from aluminum (EN-573-3), which ensures stability and lightweight. The sample holder can be moved in the X-Y plane via two linear motorized stages with built-in controller (Zaber Technologies Inc., T-LSM50A-S) assembled orthogonal to each other. We chose stepper motor-based stages as they provide a precision of a tenth of micrometer, a large travel range, have relatively low cost and comprise integrated control units with minimal footprint. They enable scanning of a 45 × 45 mm$^2$ area, which is suitable for microscope slides and Petri dishes, with a precision of ±15 $\mu$m and travel speeds ranging from 0.2 $\mu$m/s to 7 mm/s (Fig. 2(d)). Travel speeds in the $\mu$m/s range are useful, for example, in performing exposures of (bio)chemicals to the surfaces for minutes and the mm/s range enables fast reactions in the order of seconds. In addition, the movement in the broad range of $\mu$m/s to mm/s of the MFP supports two modes of operation: a continuous movement, which can create continuous lines along a surface, and a stop-and-go movement, which is used to make spots, for example, to create microarrays. The horizontal alignment of the sample can be set via three screws on the outer edges of the sample holder.[1,18] The head holder is mounted on a shorter motorized stage (Zaber Technologies Inc., T-LSM25A), having a stroke of 12.5 mm in the Z direction. The integration of X, Y and Z stages make the cMFP independent from external motorized stages (e.g. microscope stages) for scanning the sample. The head holder allows adjustments along five degrees of freedom, which is implemented by a two axis manual linear stage (Newport GmbH, MT Compact Dovetail Linear Stage 0.375 inch) providing a ±10 mm travel range. This is mounted to a tilt platform (I.L.E.E. AG, Mini-Prism Table), which provides ±3° along the X-, Y- and Z-axis (Fig. 2(a,c)). Manual adjustments are required to precisely align the head to the microscope objective and also to compensate for any irregularities of the apex, which may arise during the polishing step in the fabrication of the heads. This adjustment stage can be detached from the main body by loosening one screw.

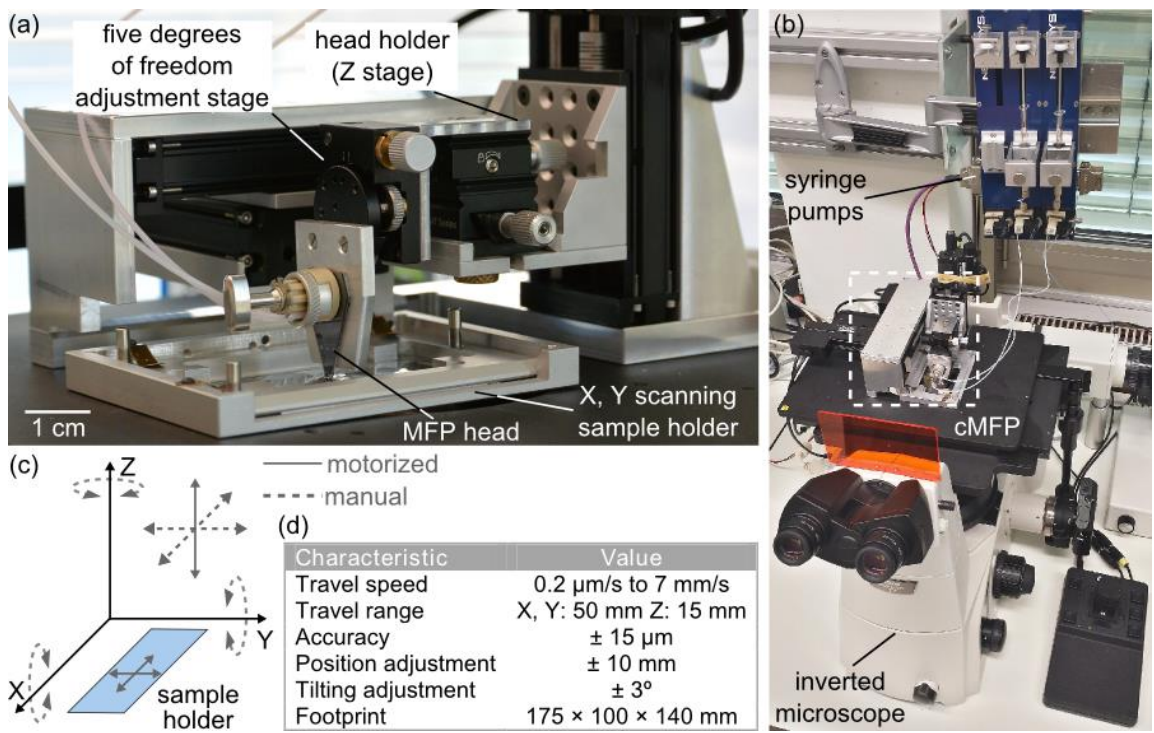

*FIG. 2.* Components and characteristics of a cMFP (a) Detailed view of the cMFP scanning system. The Z-stage (head holder) controls the distance between the head and the sample and the X-Y stage (sample holder) enables scanning $45 \times 45$ mm$^2$ of the sample of interest. (b) A cMFP placed on the stage of a standard inverted microscope. (c) Various degrees of freedom are implemented on the system. (d) Technical characteristics of the cMFP.

## C. Fluid handling

For HFC, flow rates in the $\mu$L/min range are needed. We used syringe pumps with linear motors and gear transmission (Cetoni GmbH, neMESYS Syringe Pump) together with glass microsyringes ranging from 50 $\mu$L to 1 mL (Hamilton, 1705 TLLX). The syringe pumps used generate low pulsation in the range of flow rates (1–20 $\mu$L/min) typical for the cMFP. The fluidic connections to the MFP head and the selector valve are made with standard 1/32 inch transparent tubes with an inner diameter of 0.007 in. and a length of 78 mm and PEEK fittings (Upchurch Scientific, NanoPorts). Up to four syringe pumps could be independently controlled.

### D. Selector valve

To expose a surface to multiple reagents sequentially, as done for example in immunohistochemistry,[16] the injection channel in the MFP must be connected to more than one syringe. For this, we used a selector valve (Vici, 6-to-1 dead-end Selector Valve). This selector valve has a dead volume of 204 nL and a switching time of approximately 150 ms. The output port of the selector valve was connected to one of the channels of the MFP head and the input ports to the outlets of two or three syringe pumps.

### E. Control software

We developed a LabView program for controlling the modules of the cMFP (positioning system, syringe pumps and selector valve). The positioning system can be controlled via four means: arrow buttons on the user interface, a joystick, by manually entering the coordinates of a position with the keyboard, or with a scripting module. Travel speed and acceleration can be set up to 7 mm/s and 1000 mm/s$^2$ respectively. The scripting module allows automating a sequence of movements by setting the target coordinates, the acceleration and the time delay between each movement. Scripts are text files (.txt) with four columns separated by a tab delimiter, each column corresponding to (from left to right): *axis (X, Y or Z), distance (in mm), acceleration (in mm/s$^2$) and time delay (in seconds)*. This structure of the script files allows the use of Matlab to generate scripts for following more complex paths (e.g. circular paths).

## III. MECHANICAL AND FLUIDIC INTERCONNECTS

The silicon-glass MFP heads are resistant to a broad range of chemicals and can be used indefinitely, unless physically damaged. Interfacing liquid to the microfluidic device, in general, remains a challenge in the community.[32] We developed simple to use "chip-to-world" and "chip-to-platform" interfaces. This was done by designing three mounting vias, arranged in a triangular configuration (Fig. 3(b)). The MFP head is fixed between the mounting plate and a fluidic connector (Dolomite, 8-ports Circular Connector), which is automatically aligned with the pins that enter two of the 8 ports of the connector (Fig. 3(a-c)). The mounting plate is then placed on the head holder. The "chip-to-world" interface comprises the six remaining ports to interface fluidic channels that can be mounted on the cMFP. If the selector valve is connected, there is the possibility to flow subsequently up to six liquids in one channel.

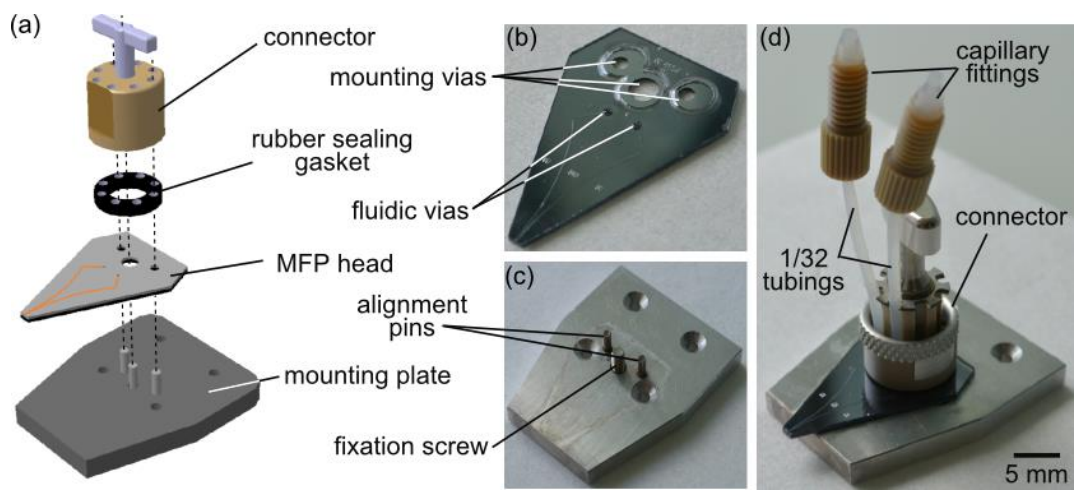

*FIG. 3. MFP heads with simple chip-to-world and chip-to-platform interfaces. (a) Exploded view of the self-alignment mechanism for mounting the cMFP head. (b) Microfabricated silicon-glass MFP head with precise through holes for mounting the head and vias for fluidic interconnection. (c) Mounting plate with alignment pins and a fixation screw. (d) The assembled interconnect system with the MFP head.*

## IV. POSITIONING SYSTEM CHARACTERIZATION

Linear stages with built-in controller constrain the motion of the MFP head in a specified direction. The motion is along straight lines, in the X, Y and the Z directions. Deviation from the trajectory or inaccuracies in the stage motion are likely due to mechanical loads, travel speed, travel range and rounding errors in the controller. We characterized the accuracy and the repeatability of the positioning system with a calibration grid patterned in chromium on a microscope slide. The position of one corner of one channel aperture of the head relative to the grid was observed with an inverted microscope and recorded using a CCD camera. A Matlab image processing toolbox was used to estimate the inaccuracy in positioning of the MFP head using the images recorded with the CCD. Accuracy is defined as the output of the system compared to the input instruction received. The procedure to characterize the accuracy of the X- and the Y-axis is: (1) zero and align the calibration grid with the MFP head, (2) move the MFP head to a known coordinate on the grid, (3) return to the starting position and again zero the position. Measurements have been performed along both axis, with travel ranges of 1, 2, 5, 10 and 15 mm for the X-axis and 1, 2, 5, 10, 15, 25, 35 and 45 mm for the Y-axis, and travel speeds of 2 and 6 mm/s (Fig. 4(a,b)). Repeatability is the measure of the variation in the position when the same input instruction is sent multiple times without zeroing the system between each repetition. The repeatability was calculated using a script that contained 10 consecutive movements going back and forth between a zero and a target position (Fig. 4(c,d)). The protocol was to perform zeroing and alignment followed by initiating a script. The same parameters as for the establishment of the accuracy were used. We note that the repeatability in the X-axis is less as compared to the Y-axis. We speculate this results from the Y-stage, which carries a heavier and unbalanced load (i.e. the center of gravity of the load is not centered on the stage) than the X-stage. Using typical operating parameters, the accuracy of the system is ±8 µm (at 6 mm/s over a 10 mm travel range). An accuracy of ±8 *µ*m is equivalent to an uncertainty of approximately one to two mammalian cells, which we believe is sufficient for many applications in the life sciences such as patterning proteins on surfaces and staining tissue sections.

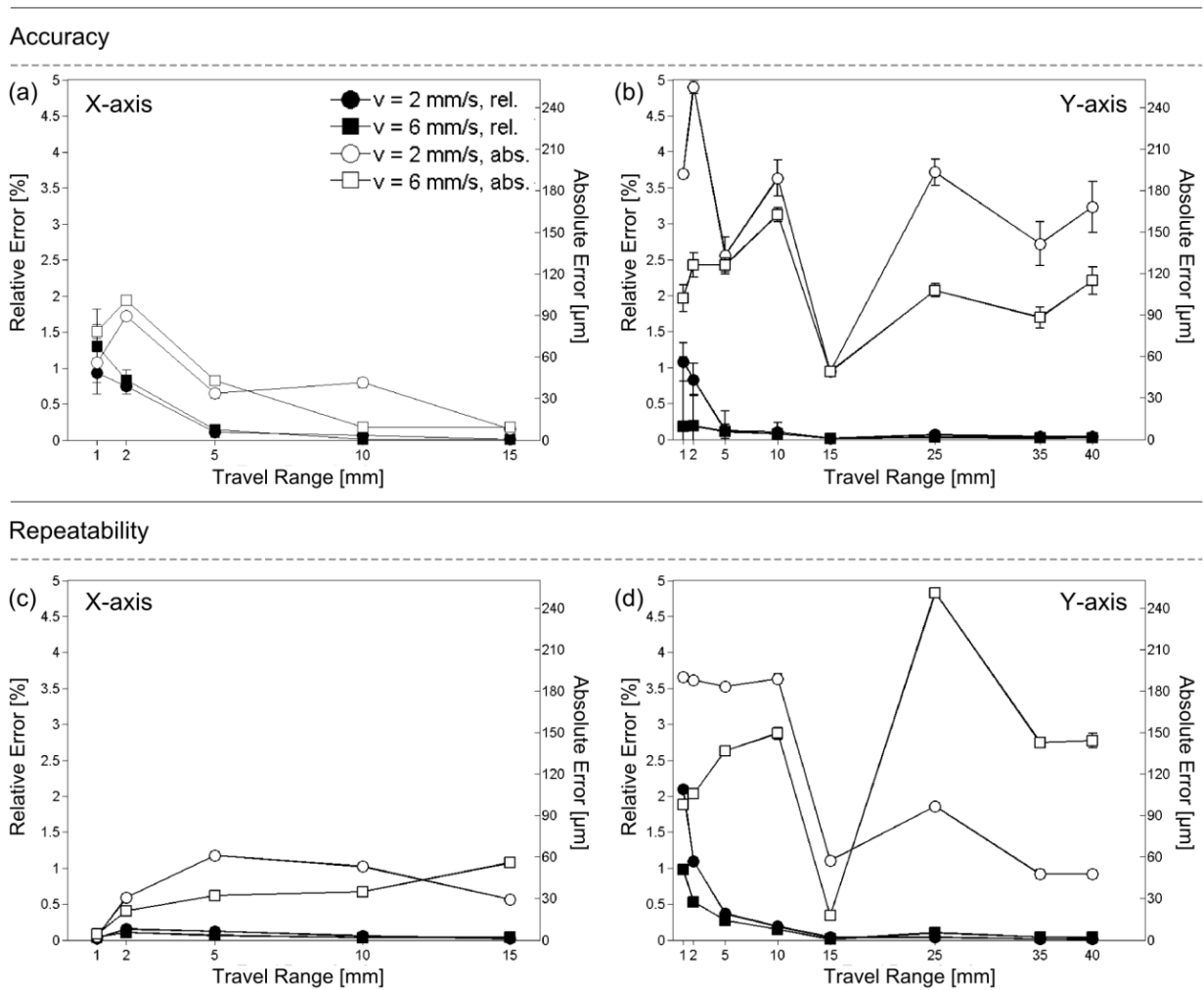

*FIG. 4. Positioning system characterization. (a–b) depict the accuracy in positioning the MFP head along the X- (a) and the Y-axis (b). The error bars represent the 95% confidence interval, obtained from 10 independent trials. (c) and (d) are plots of the repeatability in positioning the MFP head along both the X- (c) and the Y-axis (d).*

## V. FLOW VISUALIZATION AND SWITCHING LIQUIDS

For local processing of a surface, and in particular for biological surfaces, it is advantageous to have an interactive and visual feedback of the confined liquid at the apex of the MFP head. For example, in processing a tissue section with the cMFP that confines a dye, a constant view of the flow confinement from the side, and simultaneously the interaction of the dye with the tissue captured from below will provide real-time progress in staining. Importantly, this enables the use of the cMFP, unlike earlier, now with opaque semiconductor surfaces.[13,33] For the side viewing (Fig. 5(a)) we used a CCD miniature microscope (Spi GmbH, TIMM 400).

Another important aspect in processing biological surfaces is exposing them sequentially to multiple chemicals. This is a prerequisite in a large number of cell and molecular biology assays such as, successive staining of a tissue section and protein patterning − we therefore implemented and tested liquid switching. We used the selector valve to switch liquids in the flow confinement in order to expose the surface sequentially to different liquids. Two syringes were filled with water and a solution of Rhodamine B and were connected to the entry ports 1 and 2 of the selector valve. When Rhodamine B was set as an input, fluorescence in the flow confinement was observed. Every 90 seconds, the liquid flow was switched and the flow confinement was recorded with a CCD camera. The flow confinement was generated with an injection of 5 $\mu$L/min and an aspiration of 20 $\mu$L/min ($Q_i/Q_a$ = ¼). We calculated the area of the flow confinement from each frame of the recorded video using ImageJ (Fig. 5(b)). We set two thresholds at 90% and 10% light intensity, which we considered the threshold where the flow confinement was "pure" with either water or Rhodamine B. For all values beyond these set-points, the intensity of the signal was set to 90% and 10%. We observed that the time to switch between liquids at the flow confinement consists of two components: $t_{dead}$ and $t_{diff}$. $t_{dead}$ is the time until the intensity of the measured signal either increases or decreases to the 90% or the 10% set-point. Two parameters have a significant influence on $t_{dead}$, the injection flow rate, and the volume of liquid within the flow path (tubing and the channels within the MFP head) between the selector output and the apex of the head. $t_{dead}$ was measured to be 40 seconds. By increasing the injection flow rate, $t_{dead}$ can be reduced. In contrast, $t_{diff}$ is the time it takes for the dead-volume to be purged out of the flow path and was found to be 30 seconds at a flow rate of 5 $\mu$L/min, which translates to a volume of 2.5 $\mu$L. During the flow of the liquid in the fluidic path, the interface between the two sequential liquids is not "sharp", due to diffusion and the parabolic flow profile, therefore an additional volume needs to be purged.

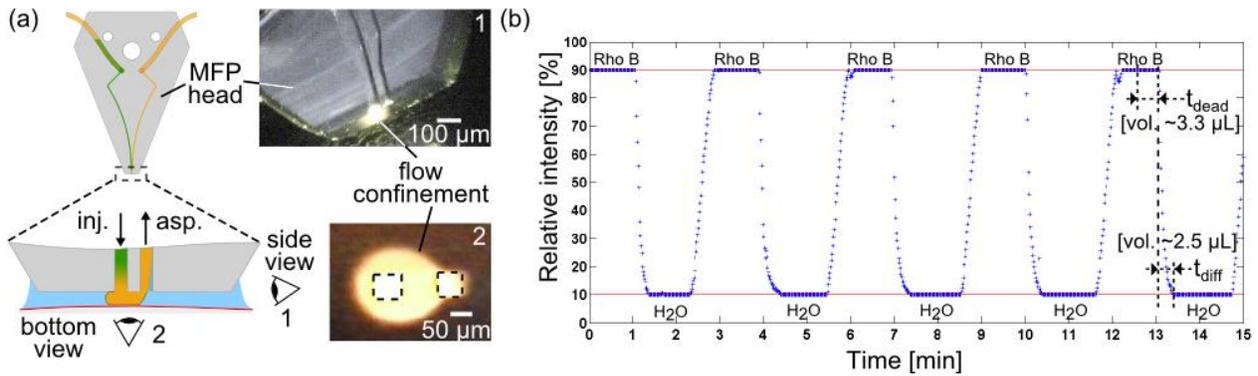

*FIG. 5. Visualization of a flow confinement and switching liquids with the cMFP. (a) Visualization of the flow confinement from the side (1) and the bottom (2) allows the cMFP to be used with transparent and opaque substrates. (b) Switching flow confinement composition between water and a solution of Rhodamine B.*

## VI. HEIGHT COMPENSATION

For tumor diagnosis, most often a 3 to 20 $\mu$m tissue section is analyzed. These sections could have topographical variation on account of the origin of the biopsy and also the subsequent processing. Such topographies are also observed within cell cultures in Petri dishes. These variations could, in general, be the equivalent of 1-3 cells translating to a few micrometers to tens of micrometers in height.

To follow the macroscopic features of irregular surfaces, we developed a height compensation algorithm that adjusts the distance of the head to the surface. The cMFP being a non-contact scanning technology, during operation a constant apex-to-surface height is required, for which height compensation within the cMFP was developed (Fig. 6(a)). The compensation applies only to the height (Z-axis), and not to the tilt of the head. The topographies that can be contoured are restricted by the dimension of the apex of the MFP heads. We developed the capability of compensating height differences up to 750 $\mu$m (data not shown). The height-compensation algorithm decomposes a profile along the surface into 1000 units. Defining a higher number of such units will result in longer scanning time, and is ultimately limited by the accuracy of the stage. The smallest step achievable by the stages used is 1 $\mu$m. For each unit of the profile along the surface, a height *h* is calculated on the basis of the topography known *a priori*. The output of the algorithm is a script of coordinates that control the positioning of the MFP head.

To demonstrate the efficacy of the height compensation of the cMFP, etched patterns on glass substrates coated with 20 nm of copper were created using the height compensation algorithm (Fig. 6(b,c)). A 20% sodium persulfate ($Na_2S_2O_8$) solution was used as etchant.[34] The substrates were a microscope slide and two plano-convex lenses with radii of 104.6 and 258.4 mm (Edmund Optics, PCX 32890 and 45280). Patterning of the Cu-coated lens, with a bend radius of 258.4 mm is shown in Fig. 6(b). We demonstrated the creation of complex patterns using the scripting module (Fig. 6(c)). The etched line patterns were within 5% variation, suggesting that the head maintains a constant distance between the apex and the lens contour (Fig. 6(b)). The dependency of the line width with the distance was shown by etching parallel lines at 30 and 50 $\mu$m from the surface. This was demonstrated in our earlier work.[1] A difference of 20 $\mu$m in the distance resulted in a 33% variation of the line width (Fig. 6(b)). We further applied this approach to irregular topographies.

The head could follow the topography of micrometer-sized structures fabricated on a microscope slide by previously acquiring the surface profile with an optical profilometer and transferring the data to the algorithm. This enabled the generation of lines with a constant width along the structure but also to pattern points inside the structures (Fig. 6(c)). Without height-compensation a dramatic change in the line width is observed.

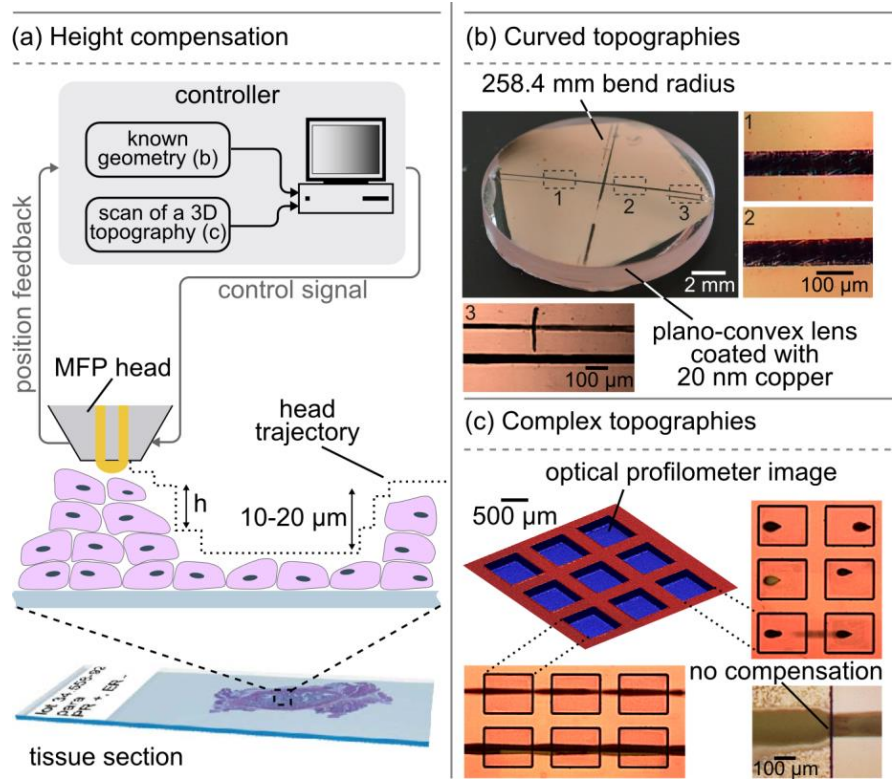

*FIG. 6. Height compensation using the cMFP. (a) A tissue section may have variations in its topography ranging from 1-3 cells (1-20 μm). For non-contact processing of tissues, to maintain a constant apex-to-surface height during scanning, the cMFP modulates h using a priori knowledge of the topography and continuous position feedback of the stages. (b) and (c) are two geometries – curved and steps – microfabricated in glass and represent the topographical variations as in tissue sections. (b) Curved surface of a Cu-coated plano-convex lens with a 258.4 mm bend radius was etched to create lines. Height compensation in the cMFP ensured the width of the lines etched along the curvature of the lens remained within 5% variation. (c) Local etching of patterns was performed on Cu-coated microfabricated structures based on the topographical information obtained a priori using an optical profilometer. Using height compensation, the line width variations were within 5% and without height compensation ~20% increase was observed.*

## VII. TISSUE STAINING

Tools that can assist in extracting more high-quality information from tissue sections by localizing the staining procedure, and that can be operated interactively provide increased flexibility in staining and are critically needed for clinical pathology, drug discovery and basic research. In particular, staining cells in a tissue section is used to visualize the level of expression of specific markers and the nuclei or cellular membrane for morphological characterization to identify the type and stage of the disease. These stained tissues are then observed under a microscope and interpretation is made by one or multiple trained pathologists.

Here, for the first time, we illustrate how a cMFP can perform controlled staining of tissue sections. To illustrate the applicability of the cMFP in pathology, we confine a solution containing hematoxylin with the cMFP and stain tissue sections. This local staining could be done using either a joystick or by entering a script, providing the interactive capability. Tissues comprise cellular heterogeneity, and different regions may require specific staining. The cMFP can make contours (Fig. 7(a)), patterns such as lines and spots (Fig. 7(b)) and also apply different chemical, at different concentrations in close physical proximity to each other (Fig. 7(c)). This not only enables to enhance the contrast of the stained region but provides an appropriate reference. In addition, by changing the head-to-surface distance $d$, the spot size of the stained tissue can be modified. Demonstration was made staining a $7 \times 7$ mm area of a cell block (Fig. 7(d)).

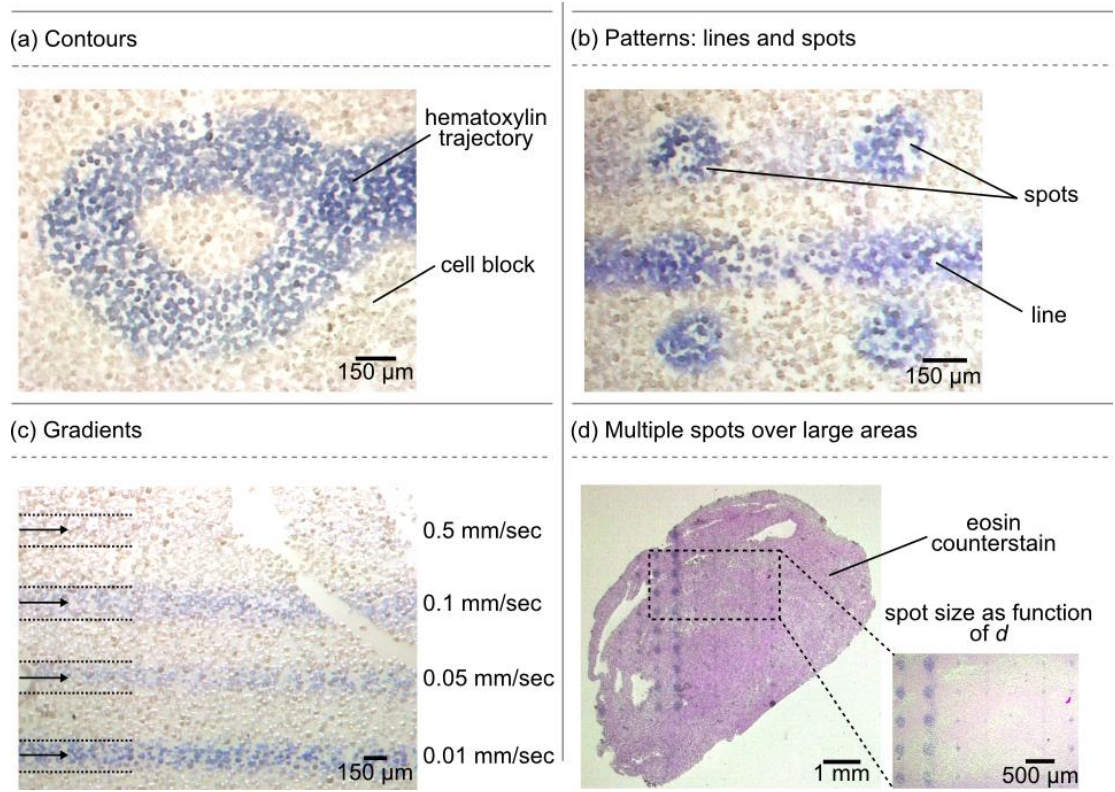

*FIG. 7. Local tissue section staining using hematoxylin for histological observation. With a joystick, in an interactive manner, contours comprising lines of hematoxylin were created (a). Continuous trajectories and spots (b) were written using the scripting module of the cMFP and gradients (c) to achieve different contrasts were obtained by varying the travel speed of the cMFP along the trajectory (arrows). In addition, a large number of spots with different sizes can be created on a tissue section (d), each row of points with different dimensions by modifying the head-to-surface distance d.*

## VIII. CONCLUDING REMARKS

The compact MFP meets the specifications for a user-friendly implementation of the MFP technology, as outlined in section II. The X-Y-Z positioning system allows scanning of standard surfaces (microscope slides, Petri dishes) with micrometer resolution. The accuracy of the positioning system is commensurate with the typical size of single cells. The small footprint of the cMFP enables it to be placed on inverted microscopes. The alignment between the MFP head, the sample and the microscope objective can be manually adjusted via the cMFP head holder. The motorized stages with stepper motors provide the flexibility of operating the cMFP in both the stop-and-go and the continuous operation mode. The implementation of standardized mechanical and fluidic interfaces for the MFP head establishes design rules to allow the user to focus on the channel configuration rather than on the interfaces.

The weight of the cMFP, not including the pumps, is about 1.8 kg. The system can be made even smaller with the use of micro electro mechanical (MEMS)-based pumps. This weight reduction along with the footprint reduction suggests that the cMFP can now be considered as a lightweight peripheral instrument, which can complement a microscope. We believe that it may even be possible to use the cMFP directly on large samples rather than bringing the sample to the instrument. Measurement instruments developed for small samples, when miniaturized, tend to reach a point of enabling them to be positioned on larger samples. This has been the case for the contact angle goniometer,[35] atomic force microscope ("beetle-type")[36] and portable X-ray fluorescence meters.[37] We further believe that for visualization, the microscope can be eliminated and a simple consumer-grade or cell phone camera can be used. It is conceivable to even use the cMFP without any visualization.

The compactness, interactive capability, and compatibility with a standard microscope and biological substrates will likely trigger the use of cMFP in broad studies and diagnostics involving cell morphology, protein marker localization in tissue, DNA studies, and tissue and cellular research in general. We believe that such a tool will profoundly impact areas in medical diagnostics, medicine and biology.

**ACKNOWLEDGMENTS**

We acknowledge financial support by the European Research Council (ERC) Starting Grant, under the 7$^{th}$ Framework Program (Project No. 311122, "BioProbe" project). We are grateful to Julien Autebert, Aditya Kashyap, Yuksel Temiz and Martina Hitzbleck for discussions. We thank Andreas Stemmer (ETH Zurich), Michel Despont, Urs Duerig and Walter Riess for their continuous support.

# REFERENCES

[1] G.V. Kaigala, R.D. Lovchik, U. Drechsler, and E. Delamarche, Langmuir **27**, 5686 (2011).
[2] G.V. Kaigala, R.D. Lovchik, and E. Delamarche, Angew. Chem. Int. Ed. **51**, 11224 (2012).
[3] R.D. Piner, J. Zhu, X. Feng, S. Hong, and C.A. Mirkin, Science **283**, 661 (1999).
[4] D.S. Ginger, H. Zhang, and C.A. Mirkin, Angew. Chem. Int. Ed. **43**, 30 (2004).
[5] L. Ying, Biochem. Soc. Trans. **37**, 702 (2009).
[6] S. Hardt and T. Hahn, Lab. Chip **12**, 434 (2012).
[7] H. Tavana, A. Jovic, B. Mosadegh, Q.Y. Lee, X. Liu, K.E. Luker, G.D. Luker, S.J. Weiss, and S. Takayama, Nat. Mater. **8**, 736 (2009).
[8] H. Tavana, B. Mosadegh, and S. Takayama, Adv. Mater. **22**, 2628 (2010).
[9] H. Tavana, B. Mosadegh, P. Zamankhan, J.B. Grotberg, and S. Takayama, Biotechnol. Bioeng. **108**, 2509 (2011).
[10] A. Ainla, G. Jeffries, and A. Jesorka, Micromachines **3**, 442 (2012).
[11] D. Juncker, H. Schmid, and E. Delamarche, Nat. Mater. **4**, 622 (2005).
[12] M.A. Qasaimeh, T. Gervais, and D. Juncker, Nat. Commun. **2**, 464 (2011).
[13] R.D. Lovchik, U. Drechsler, and E. Delamarche, J. Micromechanics Microengineering **19**, 115006 (2009).
[14] M.A. Qasaimeh, S.G. Ricoult, and D. Juncker, Lab. Chip **13**, 40 (2013).
[15] A. Queval, N.R. Ghattamaneni, C.M. Perrault, R. Gill, M. Mirzaei, R.A. McKinney, and D. Juncker, Lab. Chip **10**, 326 (2010).
[16] R.D. Lovchik, G.V. Kaigala, M. Georgiadis, and E. Delamarche, Lab. Chip **12**, 1040 (2012).
[17] H. Shiku, T. Yamakawa, Y. Nashimoto, Y. Takahashi, Y. Torisawa, T. Yasukawa, T. Ito-Sasaki, M. Yokoo, H. Abe, H. Kambara, and T. Matsue, Anal. Biochem. **385**, 138 (2009).
[18] C.M. Perrault, M.A. Qasaimeh, T. Brastaviceanu, K. Anderson, Y. Kabakibo, and D. Juncker, Rev. Sci. Instrum. **81**, 115107 (2010).
[19] A. Ainla, G.D.M. Jeffries, R. Brune, O. Orwar, and A. Jesorka, Lab. Chip **12**, 1255 (2012).
[20] A. Ainla, E.T. Jansson, N. Stepanyants, O. Orwar, and A. Jesorka, Anal. Chem. **82**, 4529 (2010).
[21] J.N. Lee, C. Park, and G.M. Whitesides, Anal. Chem. **75**, 6544 (2003).
[22] K.V. Christ and K.T. Turner, Lab. Chip **11**, 1491 (2011).
[23] G. Binnig and H. Rohrer, Angew. Chem. Int. Ed. Engl. **26**, 606 (1987).
[24] A. Bruckbauer, L. Ying, A.M. Rothery, D. Zhou, A.I. Shevchuk, C. Abell, Y.E. Korchev, and D. Klenerman, J. Am. Chem. Soc. **124**, 8810 (2002).
[25] A. Bruckbauer, P. James, D. Zhou, J.W. Yoon, D. Excell, Y. Korchev, R. Jones, and D. Klenerman, Biophys. J. **93**, 3120 (2007).
[26] P. Novak, C. Li, A.I. Shevchuk, R. Stepanyan, M. Caldwell, S. Hughes, T.G. Smart, J. Gorelik, V.P. Ostanin, M.J. Lab, G.W.J. Moss, G.I. Frolenkov, D. Klenerman, and Y.E. Korchev, Nat. Methods **6**, 279 (2009).
[27] R.D. Lovchik, G.V. Kaigala, and E. Delamarche, in *Proc. MicroTAS* (Okinawa, 2012), pp. 1444–1446.
[28] M.S. Kim, T. Kim, S.-Y. Kong, S. Kwon, C.Y. Bae, J. Choi, C.H. Kim, E.S. Lee, and J.-K. Park, PLoS ONE **5**, e10441 (2010).
[29] A.J. Blake, T.M. Pearce, N.S. Rao, S.M. Johnson, and J.C. Williams, Lab. Chip **7**, 842 (2007).
[30] G.M. Whitesides, Nature **442**, 368 (2006).
[31] S. Haeberle and R. Zengerle, Lab. Chip **7**, 1094 (2007).

[32] J. Liu, C. Hansen, and S.R. Quake, Anal. Chem. **75**, 4718 (2003).
[33] D.A. Routenberg and M.A. Reed, Lab. Chip **10**, 123 (2010).
[34] M. Abdelgawad and A.R. Wheeler, Adv. Mater. **19**, 133 (2007).
[35] Surface Analyzer™ Goniometer, Plasmatreat, http://www.plasmatreat.com/surface-determination/contact-angle-measurement-goniometer.html (last viewed: December 2, 2013).
[36] B. Gasser, A. Menck, H. Brune, and K. Kern, Rev. Sci. Instrum. **67**, 1925 (1996).
[37] P.J. Potts, in *Portable X-Ray Fluoresc. Spectrom.*, edited by P.J. Potts and M. West (Royal Society of Chemistry, Cambridge, 2008), pp. 1–12.